\documentclass[11pt]{article}

\usepackage[a4paper, top=2.5cm,bottom=2.5cm,left=2.5cm,right=2.5cm]{geometry}
\usepackage{amsmath, amstext, amsbsy, amsopn, amscd, amsfonts, amssymb}
\usepackage{graphicx,caption,subcaption}
\usepackage{booktabs}
\usepackage{setspace}
\usepackage[colorlinks=true, allcolors=blue]{hyperref}
\usepackage{natbib}
\usepackage{siunitx}
\setcitestyle{authoryear}

\newcommand{\Ro}{R_{\mathrm{o}}}
\newcommand{\Ree}{R_{\mathrm{e}}}          

\newcommand{\lo}{\lambda_{\mathrm{o}}}
\newcommand{\ro}{r_{\mathrm{o}}}            
\newcommand{\reo}{r_{\mathrm{e}}^{0}}
\newcommand{\rz}{r^{0}}                     
\newcommand{\zeo}{\zeta^{0}}
\newcommand{\er}{\hat{\mathbf{e}}_{r}}
\newcommand{\ez}{\hat{\mathbf{e}}_{z}}
\newcommand{\ero}{\hat{\mathbf{e}}_{r}^{0}}

\newcommand{\re}{r_{\mathrm{e}}}
\newcommand{\lelec}{\lambda_{\mathrm{e}}}   
\newcommand{\lr}{\lambda_{r}}
\newcommand{\lth}{\lambda_{\theta}}
\newcommand{\lz}{\lambda_{z}}

\newcommand{\Act}{\Delta}                   
\newcommand{\Actf}{\Delta_{\mathrm{f}}}     
\newcommand{\lf}{\lambda_{\mathrm{f}}}      
\newcommand{\ur}{u_{r}}                    
\newcommand{\uth}{u_{\theta}}

\newcommand{\Ef}{E^{\mathrm{f}}}            
\newcommand{\Ebd}{E^{\mathrm{bd}}}
\newcommand{\perm}{\varepsilon}             
\newcommand{\shear}{\mu}

\newcommand{\ltau}{\lambda_{\mathrm{o}}^{\tau}}  

\newcommand{\VHB}{VHB~4910}
\newcommand{\mm}{\,\mathrm{mm}}

\title{Deformation of a circular dielectric elastomer actuator:\\
experiments and a finite-elasticity approximation\\
}

\author{
Idan Z.~Friedberg,
Dar Richin,
Aviv Reuveny,
Gal deBotton\\[0.4em]
\normalsize Department of Mechanical Engineering,\\
Ben-Gurion University of the Negev, Beer-Sheva, Israel
}

\date{}

\begin{document}
\maketitle

\begin{abstract}
The deformation of a circular dielectric elastomer actuator is studied analytically and experimentally. In the limit of large prestretch and loss of tension in the actuated disk, the response of the circumferential passive annulus coincides with a previously published solution of a plane-stress problem of a membrane with a hole. This provides an estimate for the electrode coverage ratio that maximizes the actuation ratio. The same solution is modified so that it can be used at finite prestretch, with the electrode stretch retained as a measured input. Circular samples were prestretched onto a rigid frame, coated with transparent hydrogel electrodes, and tested under increasing voltage while full-field displacements were recorded by digital image correlation. The measured displacement is essentially radial, and the deformation inside the coated disk is nearly homogeneous. The electrode stretch extracted from the inner slope, inserted into the finite-prestretch estimate, reproduces the measured radial displacement throughout the passive ring. The agreement with the ideal-dielectric description of the disk then yields an upper bound on the ratio of the shear modulus to the relative permittivity, restricting the admissible range of either parameter once the other is known.
\end{abstract}

\section{Introduction}
\label{sec:intro}

A dielectric elastomer actuator is a thin elastomer film coated with compliant electrodes.
An electric field through the thickness produces a Maxwell stress that thins the film and expands it in plane \citep{pelr&etal00scie,suo10amss}.
Acrylic VHB tapes, and VHB 4910 in particular, have been the working material for much of this development: they are easy to prestretch, sustain large strain, and are available at a controlled thickness \citep{mads&etal16macrorapidcom,hong&jian26jmatsci}.
The same features make the electromechanical response sensitive to prestretch, hold time, and the choice of elastic and dielectric parameters \citep{wiss&mazz07saap,nara&atel25advengmat}.

\begin{figure}[t]
  \centering
     \includegraphics[width=0.75\textwidth]{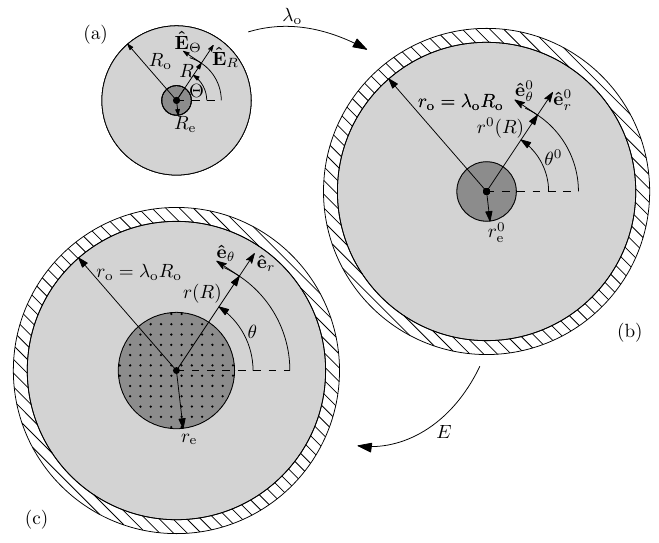}
  \caption{Planar circular EAP actuator in the (a)~reference, (b)~prestretched and (c)~electrically actuated configurations.}
  \label{fig:configs}
\end{figure}

The circular, or dot, actuator is the simplest arrangement in which that response can be demonstrated and measured.
A circular membrane is prestretched and fixed to a rigid frame at the circumference; circular soft electrodes are attached or painted at the center of the membrane, leaving a passive annulus between the coated region and the frame.
When electric potential is induced on the electrodes, the active region between the circular electrodes expands while the external annulus shrinks (see Fig.~\ref{fig:configs}).
\citet{pelr&etal00scie} used this geometry to demonstrate large voltage-induced strain.
It has since served both as a device and as a test of constitutive models.
\citet{wiss&mazz07saap} combined an analytical treatment of a homogeneously deformed coated patch with finite-element calculations of the full disk.
\citet{koh&etal11jpsppp} analyzed the deformation of dot actuators up to loss of tension point, showing that prestretch and a limited chain extensibility can suppress electromechanical instabilities.
Recently, \citet{koen&lieb24micmac} demonstrated, on the basis of a linear estimate and experimental validation, that the radial displacement is maximized when the electrode covers half the frame radius.
The same circular geometry has been used by \cite{cool&lowe22ejma} to study large-amplitude in-plane nonlinear vibrations under fluctuating voltage, including upper-branch responses with dynamic stretches of hundreds of percent, and by \cite{cool&etal26saap} to show that the geometry of the electrode runner through the passive region can produce a strong apparent frequency dependence in the measured permittivity that is not intrinsic to the elastomer.
Reviews of the field place this geometry among the standard planar actuators \citep{hong&jian26jmatsci,mads&etal16macrorapidcom}.

As the coated active disk expands, the annulus around it deforms non-uniformly and shrinks radially (the deformation from b to c in Fig.~\ref{fig:configs}).
This leads to a decrease in the radial stress in the annulus, and due to the radial stress continuity, also to a decrease in the inplane stresses in the active region.
At large enough electric field the stresses in the active region vanish.
In this loss-of-tension limit, the surrounding annulus state is of an equibiaxially stretched membrane with a traction-free hole -- a problem that was studied recently by \citep{frie&debo24mms}.
Herein we use this solution to estimate the electroelastic deformation of the circular actuator.
The same limit provides the radius for the electrodes that optimizes the coverage ratio of the circular actuator \citep{frie21msc}.

The rest of the work is structured as follows:
Section~\ref{sec:analysis} recalls the loss-of-tension result and rewrites the annulus field in the prestretched coordinates to enable comparison with the measurements, with the electrode stretch left as a measured input.
Section~\ref{sec:exp} describes the experimental setup including the circular VHB 4910 specimens, the transparent hydrogel electrodes, and the transformation of the DIC fields to polar components.
Section~\ref{sec:results} compares the measured radial displacement with the corresponding analytical approximation and, from the drop in radial stress, extracts a bound on the ratio of shear modulus to dielectric constant. A short conclusion of this work is given in Section~\ref{sec:conclusions}.

\section{Analysis of the circular actuator}
\label{sec:analysis}

The analysis follows the finite-elasticity treatment of the circular actuator in \cite{frie21msc}, built on the plane-stress solution for an equibiaxially stretched membrane with a hole~\citep{frie&debo24mms}.
The membrane is modeled as an incompressible neo-Hookean ideal dielectric~\citep{dorf&ogde14book}.
Two cases are distinguished.
First, the loss-of-tension limit, in which the radial Cauchy stress in the coated disk vanishes and the passive annulus coincides with the traction-free holed membrane, yields a closed-form electrode stretch and an optimal coverage ratio.
Second, because the present experiments do not reach that limit, the same infinite-stretch \emph{shape} of the annulus field is rewritten in prestretched coordinates with the electrode stretch \(\lelec\) prescribed by the measurement.


Three configurations are used (Fig.~\ref{fig:configs}).
In the reference configuration the membrane is a disk of outer radius \(\Ro\) with a concentric electrode of radius \(\Ree\).
An equibiaxial prestretch \(\lo\) takes the outer edge to \(\ro=\lo\Ro\), where it is fixed to the frame; the electrode moves to \(\reo=\lo\Ree\).
Application of an electric field \(E\) through the thickness of the coated disk produces a further, inhomogeneous in-plane deformation.
Material points at reference radius \(R\) occupy
\begin{equation}
  \mathbf{x}^{0} = \rz(R)\,\ero + Z\,\zeo(R)\,\hat{\mathbf{e}}_{z}^{0},
  \qquad
  \mathbf{x} = r(R)\,\er + Z\,\zeta(R)\,\ez
  \label{eq:maps}
\end{equation}
in the prestretched and actuated states, respectively.
Axisymmetry is assumed throughout.
The in-plane stretches from the reference configuration are \(\lr=r_{,R}\) and \(\lth=r/R\); incompressibility and plane stress give \(\lz=R/(r\,r_{,R})\).
The quantity to be measured in the experiments is the stretch of the electrode from the \emph{prestretched} state,
\begin{equation}
  \lelec = \frac{\re}{\reo},
  \label{eq:lelec}
\end{equation}
together with the radial displacement $\ur$.
A convenient scalar measure of visible actuation, used in the loss-of-tension calculation below is the \emph{actuation ratio}
\begin{equation}
  \Act = \frac{\re-\reo}{\ro}.
  \label{eq:Delta}
\end{equation}

The prestretched state of the whole membrane, and the actuated state of the coated disk \(R\le\Ree\), are taken as homogeneous and equibiaxial~\citep{frie21msc}:
\begin{equation}
  \rz(R)=\lo R, \qquad
  r(R\le\Ree)=\lelec\,\rz=\lelec\,\lo R,
  \label{eq:homog}
\end{equation}
the product \(\lelec\lo\) is the total in-plane stretch of the active disk.
Fringing of the electric field and of the thickness stretch at the electrode edge is neglected.


For an ideal dielectric the in-plane Cauchy stress in the coated region
\begin{equation}
  \sigma_{rr}=\sigma_{\theta\theta}
  =\shear\bigl(\lelec^{2}\lo^{2}-\lelec^{-4}\lo^{-4}\bigr)-\perm E^{2},
  \label{eq:sigactive}
\end{equation}
is written here relative to the reference configuration, where $E$ is the electric field, $\shear$ is the shear modulus, $\perm=\perm_{r}\perm_{0}$ is the permittivity with the vacuum permittivity $\perm_{0}=8.85\times10^{-12}~\text{F/m}$ and the dimensionless dielectric constant $\perm_{r}$.
As \(E\) increases, $\lelec$ increases, the radial stretch in the narrowing annulus falls, and with it the radial stress there. Continuity of radial traction at the electrode edge, together with homogeneity of the disk, implies that throughout the coated region the radial tensile stress falls beneath its value in the prestretch state ($\sigma_{rr}|_{\lelec=1}$).
The membrane cannot sustain in-plane compression, so the largest electrode radius is attained when \(\sigma_{rr}=0\).
At that state the radial traction transmitted to the annulus vanishes, and the passive region \(R\ge\Ree\) is identical to an equibiaxially stretched neo-Hookean membrane with a traction-free hole of radius \(\Ree\)~\citep{frie21msc}.

In the limit of infinite outer stretch the placement in the annulus admits the closed form~\citep{frie&debo24mms}
\begin{equation}
  \lim_{\lo\to\infty}\frac{r}{\ro}\Big|_{R\ge\Ree}
  \equiv\frac{\lambda^{\infty}(\rz)\rz}{\ro}
  =\frac{R/\Ree+\Ree/R}{\Ro/\Ree+\Ree/\Ro}
  =\frac{\rz/\reo+\reo/\rz}{\ro/\reo+\reo/\ro},
  \label{eq:rinf}
\end{equation}
where $\lambda^{\infty}$ is the stretch ratio relative to the prestretched state.
Matching~\eqref{eq:rinf} to the homogeneous disk at \(R=\Ree\) gives the electrode stretch
\begin{equation}
  \lf = \lo\,\frac{2}{1+(\Ree/\Ro)^{2}}
      = \lo\,\frac{2}{1+(\reo/\ro)^{2}},
  \label{eq:lf}
\end{equation}
and the actuation ratio
\begin{equation}
  \Actf
  =\frac{1-(\reo/\ro)^{2}}{1+(\reo/\ro)^{2}}\,\frac{\reo}{\ro}.
  \label{eq:Df}
\end{equation}
\(\Actf\) vanishes at \(\reo\to 0\) and at \(\reo\to\ro\).
Its maximum with respect to the coverage ratio is
\begin{equation}
  \left(\frac{\reo}{\ro}\right)_{\mathrm{opt}}
  =\sqrt{5}-2\approx 0.49,
  \qquad
  \Act_{\max}
  =\frac{\sqrt{5}-1}{2}\,(\sqrt{5}-2)\approx 0.30,
  \label{eq:opt}
\end{equation}
provided \(\lo\) is large enough that~\eqref{eq:rinf} applies.
This result agrees with the experimentally verified findings of \cite{koen&lieb24micmac}.
We further note in the holed-membrane problem the infinite stretch solution is already accurate for \(\lo\gtrsim 2.5\), and the intermediate-to-large transition at the optimal coverage is \(\ltau\approx 2.1\)~\citep{frie&debo24mms}.
The corresponding field \(\Ef\) follows from~\eqref{eq:sigactive} with \(\sigma_{rr}=0\).
If the dielectric strength \(\Ebd\) lies below \(\Ef\), the same \(\Act\) can still be optimized under the cap \(E\le\Ebd\); that constrained problem is not needed for the tests below.


The experiments of Sec.~\ref{sec:exp} do not reach vanishing radial stress in the disk.
The inner condition on the annulus is then a prescribed electrode stretch \(\lelec=\re/\reo\), not a free hole, while the outer edge remains fixed, \(r(\ro)=\ro\).

The stretch field $\lambda^{\infty}(\rz)$ from~\eqref{eq:rinf} is linear in \((\reo/\rz)^{2}\).
Retaining that shape and enforcing the two kinematic conditions on the prestretched configuration yields the hoop strain from the prestretched state,
\begin{equation}
  \lambda(\rz)-1
  =(\lelec-1)\frac{\lambda^{\infty}(\rz)-1}{\lambda^{\infty}(\reo)-1}
  =(\lelec-1)\frac{(\reo/\rz)^{2}-(\reo/\ro)^{2}}
        {1-(\reo/\ro)^{2}},
  \label{eq:hoopstrain}
\end{equation}
and the radial displacement
\begin{equation}
  \ur(\rz)
  =\rz\left(\lambda(\rz)-1\right).
  \label{eq:urapprox}
\end{equation}
At the electrode edge~\eqref{eq:hoopstrain} reduces to \(\lelec-1\); at the frame it vanishes.
Equation~\eqref{eq:hoopstrain} is the expression compared with the measured annulus field.
The electrode stretch \(\lelec\) will be determined from the slope of the DIC-measured radial displacements at \(\rz<\reo\).


\begin{figure}[t]
  \centering
   \includegraphics[width=0.75\textwidth]{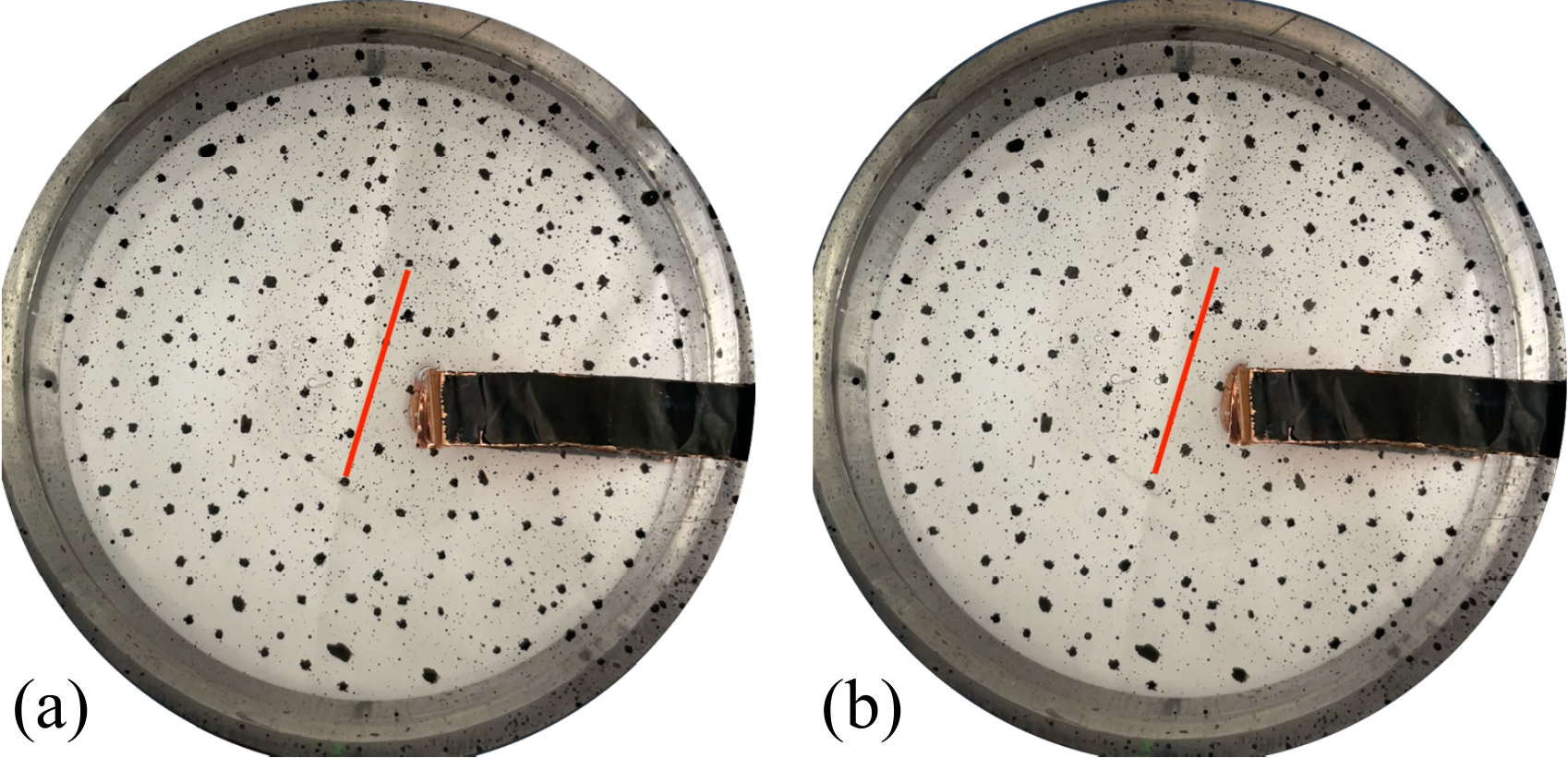}
  \caption{Experimental sample:
   (a)~Prestretched \VHB\ membrane with hydrogel electrodes and speckle pattern.
   (b)~The deformed membrane at 7000 V.
   The red segments are drawn to help visualization of the motion of the speckle.}
  \label{fig:setup}
\end{figure}

\section{Experimental setup and procedure}
\label{sec:exp}

The actuator is a thin, equibiaxially prestretched circular membrane of the acrylic elastomer \VHB, coated over a concentric disk with a pair of transparent compliant electrodes and held at its outer edge by a rigid frame.
The tests reported later are restricted to this axisymmetric geometry.

\subsection{Sample preparation}
\label{sec:specimen}

Square sheets of \VHB\ with side length \(45\mm\) were stretched equibiaxially and bonded to a square frame of outer side \(80\mm\).
The corresponding prestretch is
$  \lo = \frac{80}{45} \approx 1.77$.
The frame has a circular opening of diameter \(60\mm\), which defines the section of the membrane which is free to deform during the experiments.
In the prestretched configuration the outer radius of that membrane is therefore
 $\ro = 30\mm = \lo\,\Ro$,
and the reference outer radius of the VHB sheet that maps onto the opening is \(\Ro=\ro/\lo\).
The referential thickness of the sheets before prestretch is $H_{0}=1\text{mm}$.

To enable measurements of the planar deformation of the membrane by application of digital image correlation (DIC), a random speckle pattern was marked on one face with an acrylic marker.
Separate verification tests were conducted to verify that the chosen marker does not alter the mechanical or the electrical properties of the \VHB\ membrane.
These tests demonstrated that the marker did not cause changes in the membrane response, dielectric strength, and even breakdown field.
Subsequently, circular electrodes of radius $\reo$ were painted on both faces of the prestretched membrane, concentric with the opening.
Since conductive carbon grease, which is opaque, would hide the speckle from the camera,
transparent ionically conductive hydrogel electrodes of poly(vinyl alcohol), sodium chloride and glycerol were therefore used.
To this end we followed the preparation of \cite{peng&etal18epj}.
The gel was applied as a thin circular patch on each face.

\subsection{Experimental procedure}
\label{sec:hv}

Electric potential was applied across the two electrodes with a high-voltage amplifier under LabVIEW control, with the specimen inside a grounded enclosure.
The output voltage was monitored throughout each run.
The voltages used in the experiments lie below the dielectric breakdown and below the loss of in-plane tension in the coated section.

Full-field in-plane displacements were obtained by DICe (Digital Image Correlation Engine)  \citep{turn15sandiareport}.
A reference image was taken after prestretch and electrode application, and before voltage.
A second image was taken after the prescribed voltage had been applied.
Shown in Fig.~\ref{fig:setup}a is the reference image and in Fig.~\ref{fig:setup}b an image of the deformed membrane.
The region of interest covered the coated disk and the surrounding passive annulus out to the frame.
Only correlation points flagged as reliable by DICe were retained.

In order to translate the Cartesian DIC output to polar coordinates the center of axisymmetric deformation had to be identified.
To this end we determined the coordinate $(x_{0},y_{0})$ that minimizes the expression
\begin{equation}
\sum_{i}\left[\left(y^{(i)} - y_{0}\right)-\frac{u_{y}^{(i)}}{u_{x}^{(i)}}\left(x^{(i)} - x_{0}\right)\right]^{2},
\label{eq:comin}
\end{equation}
where $(x^{(i)},y^{(i)})$ are the coordinates of the DIC grid in the prestretched state and $(u_{x}^{(i)},u_{y}^{(i)})$ are the corresponding Cartesian displacement components.
Points at which the displacement magnitude is smaller than a tenth of the mean displacement magnitude were filtered out.
The quadratic optimization problem (\ref{eq:comin}) can be solved analytically.
The DIC results were rewritten in polar components, where
$\rz=\sqrt{(x-x_{0})^{2}+(y-y_{0})^{2}}$ and
$\theta=\arctan{(y-y_{0})/(x-x_{0})}$.
The displacements from the prestretched configuration are
$\ur = u_{x}\cos\theta + u_{y}\sin\theta$ and $\uth = -u_{x}\sin\theta + u_{y}\cos\theta$.
In Sec.~\ref{sec:results} it is demonstrated that indeed the measured \(\uth\) are negligible compared with \(\ur\).


\begin{figure}[t]
  \centering
     \includegraphics[width=0.5\textwidth]{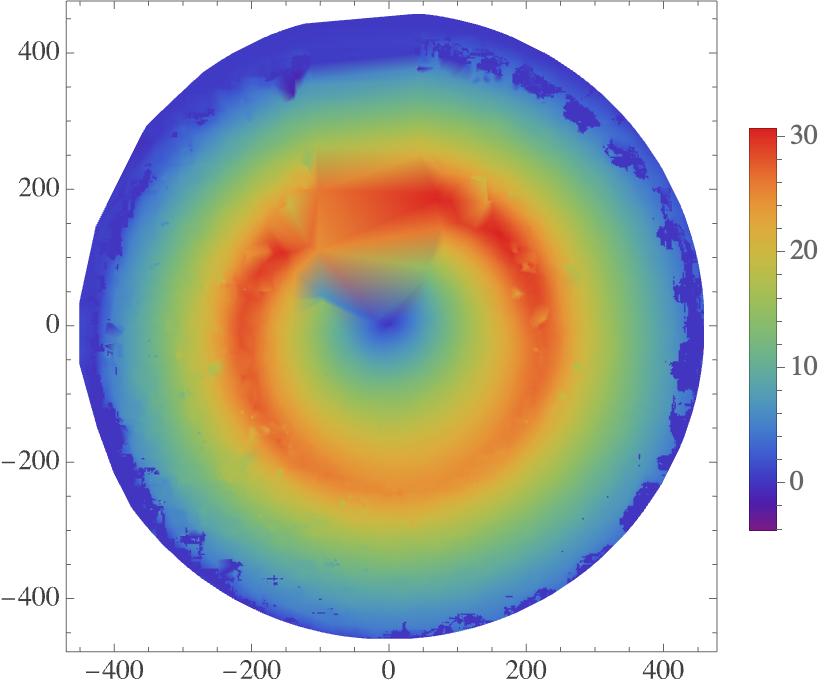}
\caption{Density plot of the DIC-measured radial displacements from electric excitation at 7000 V relative to the prestretched configuration. The coordinates and the displacements range are presented in terms of DIC pixels.}
  \label{fig:DICUrDensity}
\end{figure}

\section{Results and comparison}
\label{sec:results}

\begin{figure}[b]
  \centering
     \includegraphics[width=0.35\textwidth]{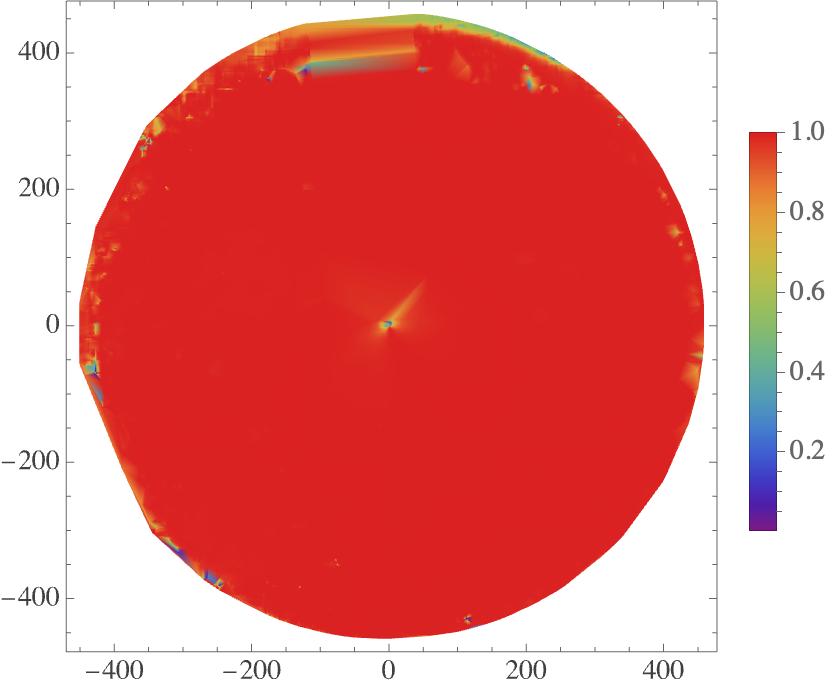}\quad\quad
     \includegraphics[width=0.35\textwidth]{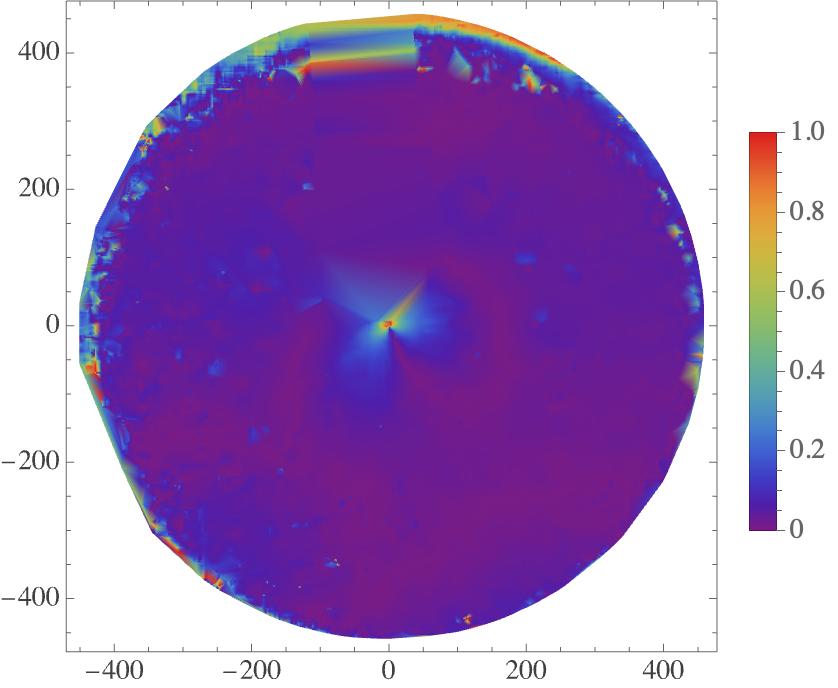}
  \caption{Density plots of the DIC-measured radial (left) and tangential (right) displacements, normalized by the displacement magnitude, from electric excitation of 7000 V relative to the prestretched configuration. Coordinates are presented in terms of DIC pixels.}
  \label{fig:DICUrqDUDensity}
\end{figure}

Experiments were carried out with electrodes whose radius is approximately $\reo\approx14\text{mm}$.
The applied voltage was increased gradually up to $V_{0}=7000~\text{V}$, and DIC snapshots were taken every 1000 V. Since the electroelastic force is proportional to $E^{2}$, meaningful deformations were measured from 4000 V. Shown in Fig.~\ref{fig:DICUrDensity} are the largest radial displacements due to electric excitation of 7000 V as measured by the DIC system relative to the prestretched configuration.
The coordinates and the displacements range are in terms of the DIC pixels.
The displacements show an axisymmetric pattern, with peak shifts of approximately 30 pixels at the electrode circumference.
Density plots of the ratios $\ur/u$ (left) and $\uth/u$ (right) are shown in Fig.~\ref{fig:DICUrqDUDensity}.
Throughout the specimen the ratio $\ur/u$ is close to unity with exceptions in small regions at the center and the periphery where the displacements are small.
The tangential displacements shown in the right plot are mostly smaller than 5\% of the total displacement.
The exceptions are, once again, near the center and in the peripheral region of the outer annulus where the displacements are small.
These further demonstrate the axial symmetry of the displacement field relative to the coordinate $(x_{0},y_{0})$.

\begin{figure}[t]
  \centering
     \includegraphics[width=0.5\textwidth]{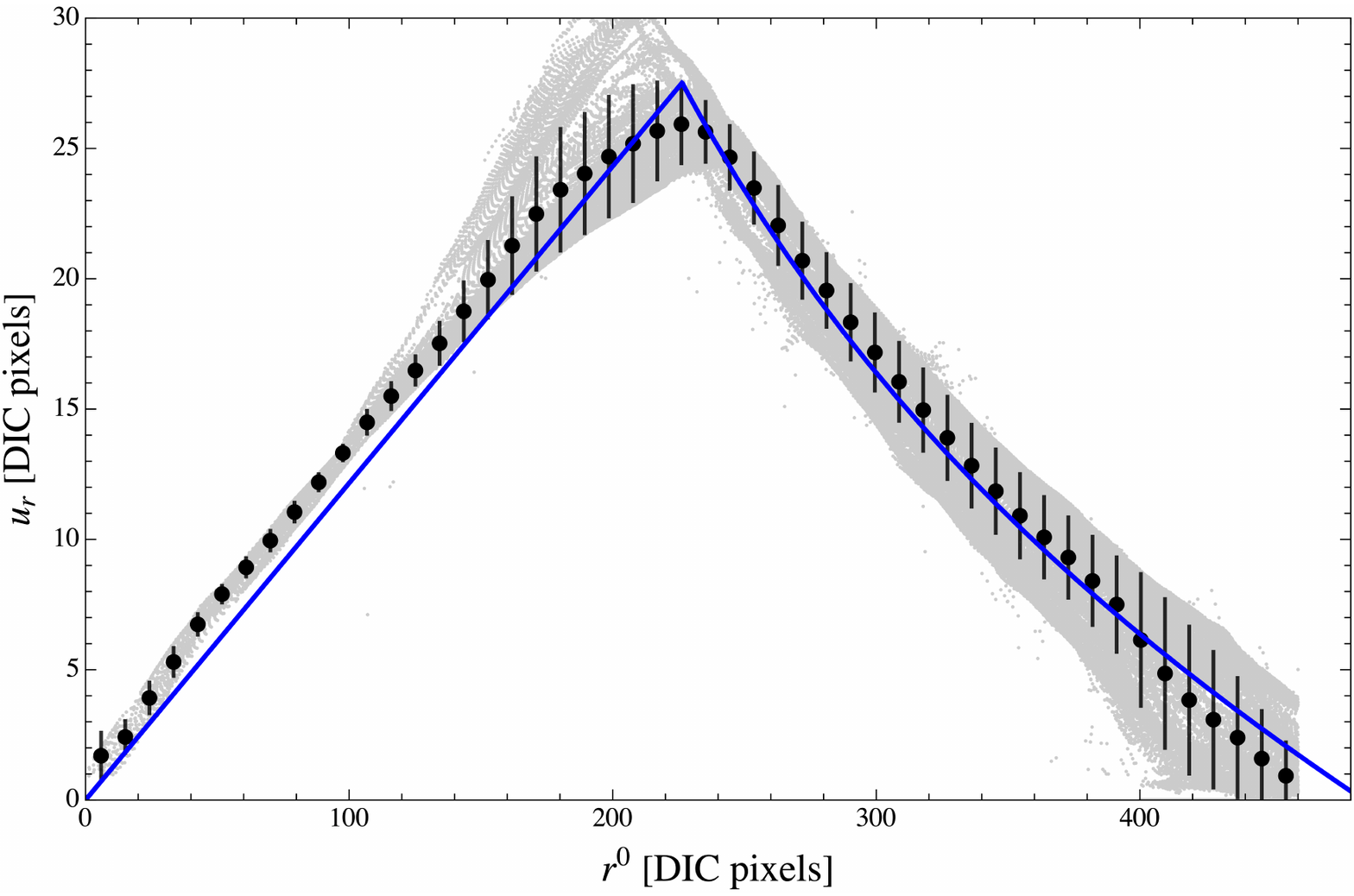}
  \caption{DIC-measured radial displacements versus the proposed approximation at $7000\,\mathrm{V}$. Gray: individual correlation points; Black: means in $N=50$ annuli; bars: one standard deviation; Blue: proposed approximation.}
  \label{fig:DICUofR7000}
\end{figure}

To obtain the mean radial displacement as a function of the prestretched radius \(\rz\), the DIC cloud was partitioned into \(N\) concentric annuli. In each annulus the mean of \(\ur\) and the standard deviation about that mean were computed. Fig.~\ref{fig:DICUofR7000} shows the raw DIC values of \(\ur\) (gray) together with the annular means for \(N=50\) (black) and the corresponding standard deviations (vertical bars) at \(7000\,\mathrm{V}\). Means obtained with \(N>20\) differ only negligibly from those with \(N=50\), so \(N=50\) will be used in the remainder of the analysis. The clouds recorded at the other voltages have the same shape as the \(7000\,\mathrm{V}\) field, scaled with the applied voltage.
Over the coated disk the scatter is modest and even near the electrode edge the standard deviation remains below 10\% of the local mean. Only in the outer passive ring, as \(\rz\to\ro\) and \(\ur\) itself becomes small, do the relative variations become large.

\begin{figure}[b]
  \centering
     \includegraphics[width=0.55\textwidth]{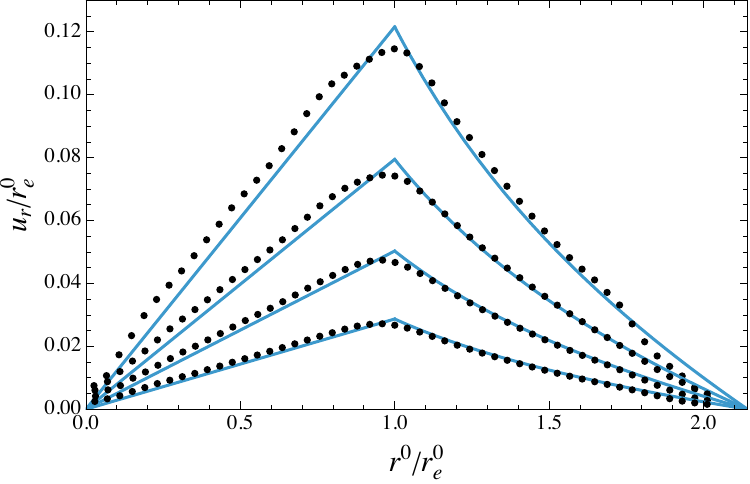}
  \caption{Mean DIC-measured radial displacements (black dots) versus the proposed approximations (blue curves) at $V_{0}=4000$, 5000, 6000 and 7000 $\mathrm{V}$ from bottom to top, respectively.}
  \label{fig:urvsr0all}
\end{figure}

The slope of the nearly linear dependency of \(\ur\) on \(\rz\) in the inner portion of the coated region was used to determine the uniform stretch ratio $\lelec$ of the actuated region.
In the region $\rz<2/3\reo$ for the 7000 V experiment the measured stretch ratio is $\lelec\approx1.121$.
Taking \(\ur=(\lelec-1)\rz\) for $\rz<\reo$ and Eq.~(\ref{eq:urapprox}) with (\ref{eq:hoopstrain}) in the outer regime, is the proposed straightforward approximation for the membrane deformation.
The blue curve in Fig.~\ref{fig:DICUofR7000} shows this approximation in comparison with the DIC-measured deformations.
In particular we observe the neat agreement of the curve with the experimental results in the external passive annulus.

Similar analyses were carried out for the rest of the DIC snapshots.
The resulting approximations are shown together the experimentally measured mean values of the radial displacements at 4000, 5000, 6000 and 7000 $\mathrm{V}$ in Fig.~\ref{fig:urvsr0all}.
The displacements and the radii are normalized by the electrodes radius in the prestretched state $\reo$.
Once again we observe the ability of Eq.~(\ref{eq:urapprox}) with (\ref{eq:hoopstrain}) to accurately predict the measured displacement in the outer section even at relatively lower hoop stretches.


\begin{figure}[t]
  \centering
  \begin{subfigure}[t]{0.47\textwidth}
    \centering
    \includegraphics[width=\textwidth]{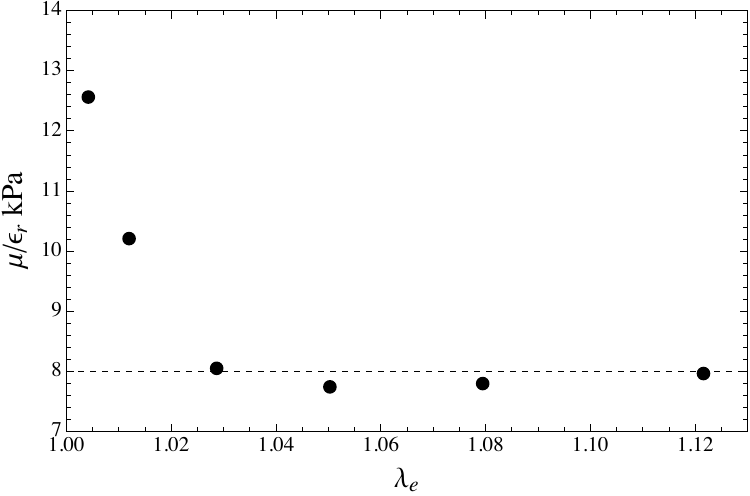}
    \caption{}
    \label{fig:bound}
  \end{subfigure}
  \hfill
  \begin{subfigure}[t]{0.47\textwidth}
    \centering
    \includegraphics[width=\textwidth]{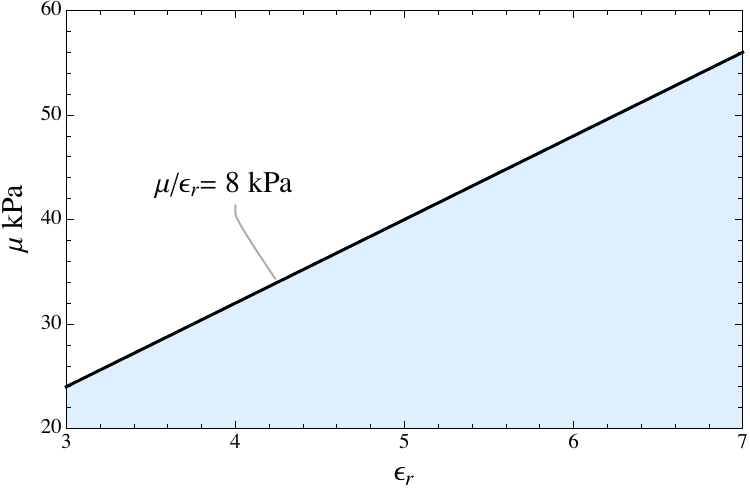}
    \caption{}
    \label{fig:mu}
  \end{subfigure}
  \caption{Constitutive parameters: (a) The bound \eqref{eq:boundSoP} on \(\shear/\perm_{r}\)
  as a function of the stretch ratio in the active region.
  (b) Acceptable range of the shear modulus (blue) for the
  \(8\,\mathrm{kPa}\) bound.}
  \label{fig:boundparameter}
\end{figure}

The proposed estimation for the displacement field does not rely on estimates for either the shear modulus or the permittivity of the membrane. Yet, to determine the stresses and the electroelastic energy these constitutive parameters are required. Specifically, the decrease in the radial stress in the active disk relative to the stress in the prestretched state is:
\begin{equation}
  \sigma_{rr}-  \sigma_{rr}|_{\lelec=1}
  =\shear\bigl(\lelec^{2}\lo^{2}-\lelec^{-4}\lo^{-4}\bigr)-\perm E^{2}-\shear\bigl(\lo^{2}-\lo^{-4}\bigr).
  \label{eq:dofsigactive}
\end{equation}
Since this is negative, we can deduce the following bound on the ratio between the shear modulus and the dielectric constant,
\begin{equation}
  \frac{\shear}{\perm_{r}}
  \le\frac{{\perm_{0}}{\lo^{4}}\lelec^{4}V_{0}^{2}}
  {{H_{0}^{2}}\bigl[(\lelec^{2}-1)\lo^{2}-(\lelec^{-4}-1)\lo^{-4}\bigr]},
  \label{eq:boundSoP}
\end{equation}
where we substitute $E=-\lo^{2}\lelec^{2}{V_{0}}/{H_{0}}$.

\begin{table}[t]
  \centering
  \caption{Electrode stretch from the slope of \(\ur(\rz)\) for \(\rz<2/3\,\reo\),
  and the corresponding upper bound on \(\shear/\perm_{r}\).}
  \label{tab:lelec}
  \begin{tabular}{c S[table-format=1.3] S[table-format=2.2]}
    \toprule
    {\(V_{0}\) (V)} & {\(\lelec\)} &  {\(\shear/\perm_{r}\) kPa} \\
    \midrule
    2000 & 1.004 &  12.56 \\
    3000 & 1.012 &  10.21\\
    4000 & 1.029 &  8.05\\
    5000 & 1.050 &  7.74\\
    6000 & 1.079 &  7.80\\
    7000 & 1.121 &  7.96 \\
    \bottomrule
  \end{tabular}
\end{table}

Based on the DIC measurements, the bound (\ref{eq:boundSoP}) is shown in Fig.~\ref{fig:boundparameter}a as a function of $\lelec$ (see also Table~\ref{tab:lelec}). At strains smaller than 0.03 ($V_{0}<4000 \text{V}$), the bound drops from about 13 to 8 kPa. At larger strains and electric excitations, its value is almost fixed at 8 kPa (the horizontal dashed line). Taking the lowest restriction on the upper bound at 8 kPa, in Fig.~\ref{fig:boundparameter}b the predicted acceptable range of values for the shear modulus (blue region) is plotted for a range of dielectric constants between 3 and 7 \citep[e.g.,][]{vhbdata3m22,kofo&etal03jimss,mcka&cali09inbook,bort&etal16ijss,cohe&etal17emlet,cool&etal26saap}.
Note that the acceptable range for the shear modulus, between 30--50 kPa, is in agreement with some published values \citep[e.g.,][]{plan&dubo06ijss,bort&etal16ijss,hu&etal19inbook}, but lower than other commonly used ones \citep[e.g.,][]{vhbdata3m22,wiss&mazz07saap,koh&etal11jpsppp,dili&etal12jap}.
We emphasize that both constitutive parameters, $\shear$ and $\perm_{r}$, depend on the stretch level and the intensity of the electric field \citep[e.g.,][]{mcka&cali09inbook,cohe&gdb16prl}.

\section{Concluding remarks}
\label{sec:conclusions}

An analytical solution for the circular dielectric elastomer actuator was developed in the limit of infinite prestretch for an incompressible neo-Hookean ideal dielectric. In that limit the passive annulus coincides with the plane-stress problem of a neo-Hookean membrane containing a traction-free hole, for which a closed-form placement is available \citep{frie&debo24mms}. Matching the annulus field to a homogeneously deformed coated disk supplies the electrode stretch at loss of tension and, from it, the coverage ratio that maximizes the actuation ratio. The same infinite-stretch shape of the annulus modified so that the estimate could be used at the finite prestretch of the experiments with the electrode stretch retained as an input.

Circular VHB 4910 membranes were prestretched onto a rigid frame, coated with transparent hydrogel electrodes over a concentric disk, and tested under increasing voltage. Full-field in-plane displacements were recorded by digital image correlation. Optimized transformation of the Cartesian measurements to polar components confirmed that the displacement field is essentially radial about the identified center and that the deformation inside the coated disk is nearly homogeneous.

The electrode stretch was extracted from the slope of the radial displacement in the homogeneous inner region and substituted into the finite-prestretch estimate. The resulting radial-displacement curve agrees well with the annular means of the DIC data throughout the passive ring. Because the measured fields are consistent with the ideal-dielectric description of the coated disk, the observed drop in radial stress relative to the prestretched state yields a simple upper bound on the ratio of the shear modulus to the relative permittivity. Once one of the two electroelastic parameters is taken from an independent measurement, the bound restricts the admissible range of the other. 

\section*{Acknowledgements}
During the preparation of this manuscript, the authors utilized Grok (xAI) to assist with literature compilation, text rephrasing, LaTeX formatting, and Mathematica programming for some of the plots preparation and visualization. The authors independently developed and verified all scientific content, analyses, results, and conclusions. The authors confirmed the authenticity and relevance of all cited references.

\begin{small}

\end{small}

\end{document}